# SHOULD ENGINEERS BE CONCERNED ABOUT VULNERERABILITY OF HIGHWAY BRIDGES TO POTENTIALLY-INDUCED SEISMIC HAZARDS?

F. Khosravikia[1], A. Potter[2], V. Prakhov[3], P. Clayton[4], and E. Williamson[5]

## ABSTRACT

This paper evaluates the vulnerability of highway bridges in areas subjected to human induced seismic hazards that are commonly associated with petroleum activities and wastewater disposal. Recently, there has been a significant growth in the rate of such earthquakes, especially in areas of Texas, Oklahoma, and Kansas. The magnitudes of these earthquakes are usually lower than tectonic earthquakes that can occur in high seismic regions; however, such induced earthquakes can occur in areas that historically have had negligible seismicity. Thus, the infrastructure in these locations was likely designed for no to low seismic demands, making them vulnerable to seismic damage. Ongoing research is aimed at evaluating the vulnerability bridge infrastructure to these human induced seismic hazards. In this paper, fragility curves are developed specifically for steel girder bridges by considering major sources of uncertainty, including uncertainty in ground motions and local soil conditions expected in the Texas, Oklahoma, and Kansas region, as well as uncertainty in design and detailing practices in the area. The results of this fragility analysis are presented herein as a basis for discussion of potential seismic risks in areas affected by induced earthquakes.

[1]PhD student, Dept. of Civil Engineering, University of Texas at Austin, Austin, TX; email: farid.khosravikia@utexas.edu
[2]Engineer, Collins Engineers, Midvale, UT; e-mail: akpotter630@gmail.com
[3]Engineer, M.S.E, E.I.T., Thornton Tomasetti, Dallas, TX; e-mail: mail@prakhov.com
[4]Assistant professor, Dept. of Civil, Architectural, and Environmental Engineering, The University of Texas at Austin, Austin, TX; e-mail: clayton@utexas.edu
[5]Professor, Dept. of Civil, Architectural, and Environmental Engineering, The University of Texas at Austin, Austin, TX; e-mail: ewilliamson@mail.utexas.edu

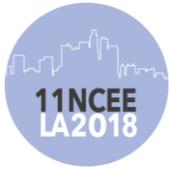

# Should Engineers be Concerned about Vulnerability of Highway Bridges to Potentially-Induced Seismic Hazards?

F. Khosravikia[1], A. Potter[2], V. Prakhov[3], P. Clayton[4], and E. Williamson[5]

## ABSTRACT

This paper evaluates the vulnerability of highway bridges in areas subjected to human induced seismic hazards that are commonly associated with petroleum activities and wastewater disposal. Recently, there has been a significant growth in the rate of such earthquakes, especially in areas of Texas, Oklahoma, and Kansas. The magnitudes of these earthquakes are usually lower than tectonic earthquakes that can occur in high seismic regions; however, such induced earthquakes can occur in areas that historically have had negligible seismicity. Thus, the infrastructure in these locations was likely designed for no to low seismic demands, making them vulnerable to seismic damage. Ongoing research is aimed at evaluating the vulnerability bridge infrastructure to these human induced seismic hazards. In this paper, fragility curves are developed specifically for steel girder bridges by considering major sources of uncertainty, including uncertainty in ground motions and local soil conditions expected in the Texas, Oklahoma, and Kansas region, as well as uncertainty in design and detailing practices in the area. The results of this fragility analysis are presented herein as a basis for discussion of potential seismic risks in areas affected by induced earthquakes.

## Introduction

Recently, there has been a significant increase in the rate of ground motions associated with petroleum activities and wastewater disposal in areas such as Texas, Oklahoma, and Kansas. Such activities generally increase the subsurface pore pressure, which facilitates the release of stored tectonic stress along an adjacent fault. These types of human-caused earthquakes generally occur in areas nearby wastewater injection wells, many of which have been historically considered as non- or low-seismic regions. Accordingly, the infrastructure around these sites has not been designed for seismic demands, raising concerns about the safety of nearby buildings and bridges. Moreover, predicting seismic hazard in areas with induced seismicity significantly changes from one year to another because of the large uncertainties in the causes of such earthquakes and the uncertainty associated with fluctuations in industrial activity. Therefore, characterizing vulnerability across such locations can be challenging. The main objective of this study is to evaluate the effect of such ground motions on bridge vulnerability to see if they are potentially damaging to bridges. To do so, fragility functions providing the conditional probability, which gives the likelihood that a structure meets or exceeds a pre-defined level of damage given ground motion intensity measures, are developed. These fragility curves will be used in ongoing research assessing seismic risk in areas affected by induced earthquakes.

[1]PhD student, Dept. of Civil, Architectural, and Environmental Engineering, The University of Texas at Austin, Austin, TX; email: farid.khosravikia@utexas.edu
[2]Engineer, Collins Engineers, Midvale, UT; e-mail: akpotter630@gmail.com
[3]Engineer, M.S.E, E.I.T., Thornton Tomasetti, Dallas, TX; e-mail: mail@prakhov.com
[4]Assistant professor, Dept. of Civil, Architectural, and Environmental Engineering, The University of Texas at Austin, Austin, TX; e-mail: clayton@utexas.edu
[5]Professor, Dept. of Civil, Architectural, and Environmental Engineering, The University of Texas at Austin, Austin, TX; e-mail: ewilliamson@mail.utexas.edu

A case study is performed considering bridges in Texas, which is one of the areas that are believed to be subjected to human-caused seismic hazards. According to Frohlich et al. [1] and Hornbach et al. [2], the rate of the potentially human-induced ground motions in Texas with a magnitude greater than 3.0 has considerably increased from approximately two per year prior to 2008 to approximately twelve per year in 2016. These authors showed that most of these earthquakes are believed to be human-caused ground motions associated with petroleum activities and wastewater disposal. For instance, Frohlich et al. [1] showed that the increase in seismicity rates correlates to the increase in hydraulic fracturing petroleum activities. According to Hincks et al. [3], the seismic moment released in Texas are strongly correlated with the waste water injection depth. Moreover, this study focuses on the seismic performance of multi-span, simply supported steel girder bridges, which are one of the common types of highway bridges. There are many studies in the literature that investigated the seismic performance of such bridges in areas with natural seismic hazards [4,5]; however, the vulnerability of such bridges in areas with potentially-induced seismic hazards is yet to be studied.

This study utilizes a probabilistic framework, which considers uncertainty in ground motions and local soil conditions, as well as uncertainty in design and detailing practices over the past several decades when the bridge population was constructed. In this framework, fragility functions are provided to evaluate the vulnerability of Texas bridges to such seismic hazards. A fragility function provides the conditional probability that gives the likelihood of whether a structure meets or exceeds a pre-defined level of damage (i.e., limit states, given ground motion intensity measures). Peak Ground Acceleration (PGA) is used as a measure for ground motion intensity. For each limit state in the proposed framework, the probability of the damage, $p_f$, is the probability that the structural demand, $D$, meets or exceeds the structural capacity, $C$, which reads:

$$p_f = P[D/C > 1 \mid \text{PGA}] \tag{1}$$

Assuming lognormal distributions for demand and capacity, Cornell et al. [6] showed that the above-mentioned probability can be computed using the following equation:

$$p_f = \Phi[\frac{\ln(S_D / S_C)}{\sqrt{\beta_{D|PGA}^2 + \beta_C^2}} \mid \text{PGA}] \tag{2}$$

where $S_D$ and $\beta_{D|PGA}$ are, respectively, the median and conditional lognormal standard deviation of the seismic demand, which will be explained later in the paper; and $S_C$ and $\beta_C$ are, respectively, the median and dispersion of the capacity. Therefore, to compute the probability of damage, models are required to estimate the median and dispersion of the seismic demands and capacities.

The main steps to produce fragility curves are shown in Figure 1. As seen, to take into account the uncertainty in the ground motion, the first step is to obtain a suite of ground motions that represent the seismicity of the studied area. Second, bridge samples are randomly selected from the bridge inventory, and each selected bridge sample is subjected to randomly selected ground motions scaled to different PGA levels. For each ground motion-bridge pair, a nonlinear time history analysis is conducted. The outputs of the nonlinear analyses, the PGA of the selected ground motion, and the demands of different components (e.g., bearings, abutments, and columns) are set as an input for the probabilistic seismic demand model (PSDM) to predict the $S_D$ and $\beta_{D|PGA}$. In addition, for each component, the probabilistic seismic capacity model (PSCM) is developed to predict $S_C$ and $\beta_C$. Having both PSDM and PSCM for each component, the fragility curves can be computed using Eq. (2). Finally, by developing the fragility curves for each component, the fragility functions for a particular bridge class can be computed. Each step of this numerical fragility procedure is described in detail in the following sections.

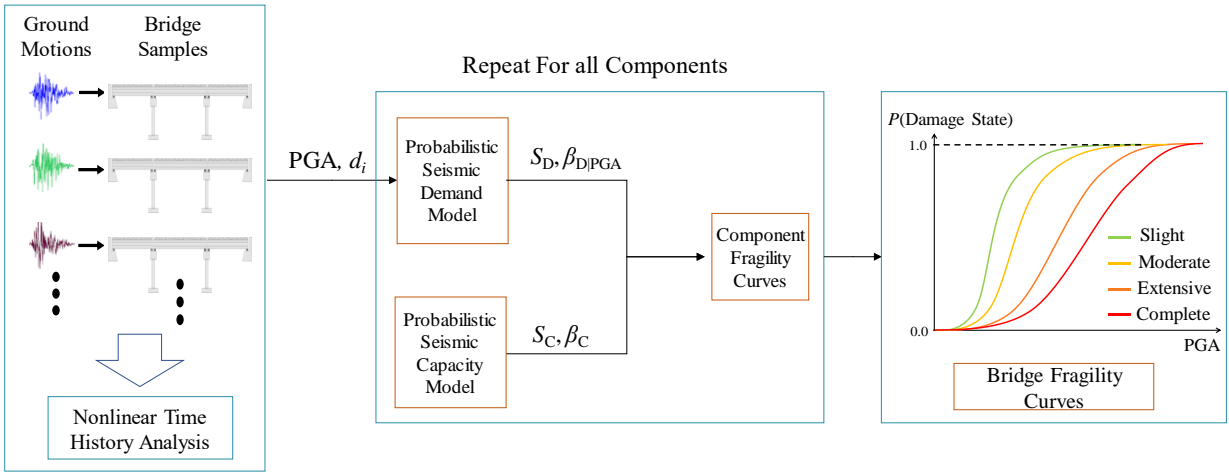

Figure 1. Analytical fragility function procedure

## Case Study

As noted, this study evaluates the vulnerability of bridges located in Texas, which is one of the areas believed to be subjected to induced seismic hazards. To do so, ground motions that represent Texas seismicity and bridge samples that represent the Texas bridge inventory are considered. In the following, the two key factors for vulnerability assessments are described in detail.

*Ground Motions*

As seen in Figure 1, the very first step for analytical fragility functions is to obtain a suite of ground motions that represent Texas seismicity. To do so, ground motion prediction models specific to the geologic and soil conditions across the state are developed to predict the intensity of ground shaking; see elsewhere for details [7–9]. To properly take into account the uncertainty in seismicity, more than 200 ground motions that have occurred since 2005, the majority of which are recorded in Texas, Oklahoma, and Arkansas, are considered. The magnitudes of these records are between 3.6 and 5.8. The depth of these ground motions varies from 2.4 km to 14.2 km, which denotes that most of them are shallow-depth ground motions. In addition, the PGA of these records goes up to 0.6 g.

Figure 2 shows the 5%-damped elastic pseudo-acceleration spectra, $S_a$, of these records normalized with respect to their PGA, together with the overall mean. This figure clearly shows the significant amount of uncertainty in the ground motion. As seen, for almost all records, $S_a$ has higher values for periods less than 0.5 second. Then, spectral accelerations rapidly diminish as the natural period increases.

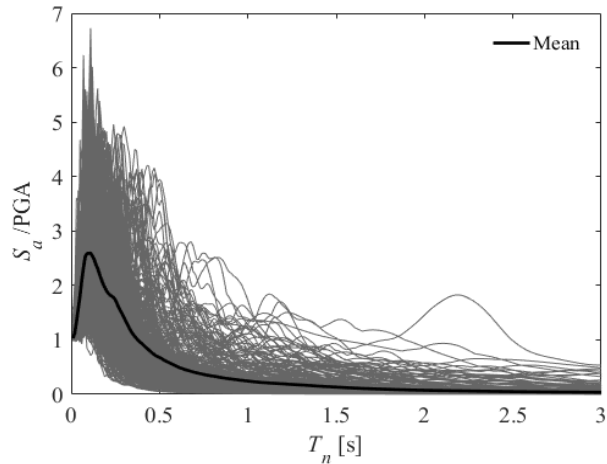

Figure 2. Normalized response spectra of ground motions in respect with their PGA

*Steel Girder Bridges*

Figure 3 shows a schematic view of a multi-span, simply supported steel girder bridge, which is hereafter referred to as steel girder bridge for brevity. Khosravikia et al. [10] showed that this type of bridge was popular in 1960s in Texas when less attention was given to structural seismic demands. The main geometric parameters of this bridge class that are critical for developing numerical bridge models to simulate seismic behavior are the number of spans, span length, vertical underclearance, and deck width. Vertical underclearance refers to the total height of the column, bearing, and bent cap, which can be used as a proxy for estimating column height for the numerical bridge models. The probability distributions of these parameters are extracted from FHWA National Bridge Inventory (NBI) [11] and Texas Department of Transportation (TxDOT) bridge database and are shown in Figure 4. As seen, this bridge class generally has two to eight spans, the lengths of which varies between 20 ft. and 90 ft. The vertical underclearance of these bridges also varies between 13 ft. and 24 ft., and their decks have widths ranging from 20 ft. to 80 ft. The average of each parameter is also shown in Figure 4.

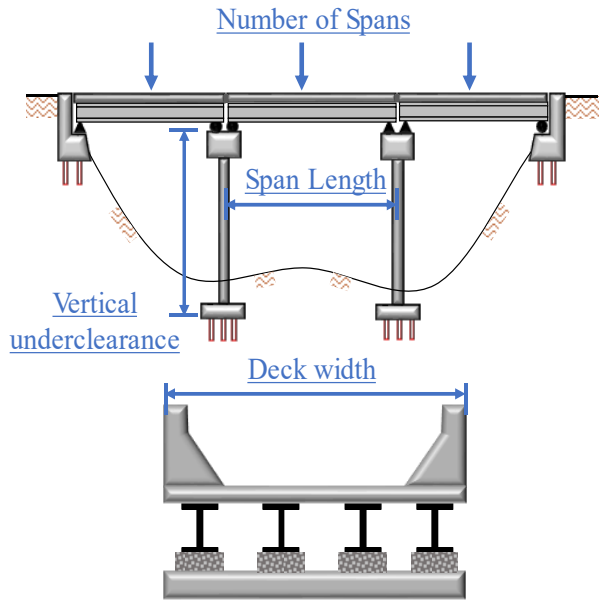
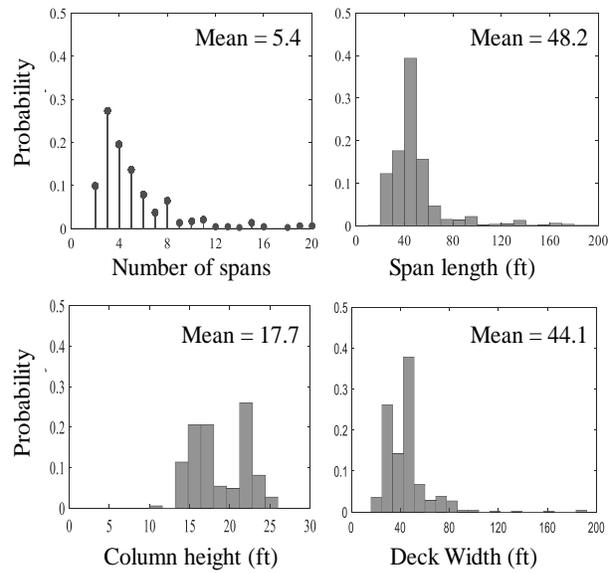

Figure 3. Schematic view of steel girder bridges

Figure 4. Geometric characteristics

According to the abovementioned distributions, eight bridge configurations are sampled from the population of the bridge class inventory to account for the variability of geometry. The sampling analysis is done using the Latin Hypercube Sampling method, taking into account the correlation among the geometric parameters. The geometric parameters of each bridge configuration are shown in Table 1. As seen in the table, the bridge configurations have three to five spans with lengths between 26.0 ft. and 77.0 ft. The uncertainty in material properties is also taken into account by considering them as random variables [12]. The range of material parameters comes from the TxDOT and NBI databases as well as other relevant studies for CEUS bridges [13]. To properly account for the effect of the uncertainty in material properties, eight different bridges with varying material properties are randomly generated for each bridge configuration using the Latin Hypercube Sampling method; see elsewhere for details [12]. The natural periods of these bridges vary between 0.25 and 0.50 seconds. In the following section, the modeling details of 64 generated bridges are presented.

Table 1. Geometric parameters of representative bridge configurations

| Bridge No. | Spans | Span Length (ft) | Deck Width (ft) | Vertical underclearance (ft) |
|---|---|---|---|---|
| 1 | 3 | 35.0 | 54.0 | 23.0 |
| 2 | 4 | 40.0 | 29.3 | 15.3 |
| 3 | 4 | 77.0 | 44.6 | 15.0 |
| 4 | 3 | 45.0 | 44.0 | 16.8 |
| 5 | 2 | 36.0 | 46.0 | 16.3 |
| 6 | 5 | 40.0 | 46.3 | 18.8 |
| 7 | 5 | 26.0 | 40.0 | 17.3 |
| 8 | 3 | 50.0 | 35.3 | 16.7 |

## Bridge Model

The behavior of the 64 sampled bridge samples are simulated in the *OpenSees* analysis program [14] using three-dimensional models. The software provides robust nonlinear dynamic analysis capabilities with numerous built-in and user-defined materials to represent a wide range of nonlinear behaviors. The developed models contain beam-column elements for the columns, bent caps, and girders, with concentrated translational and/or rotational springs to simulate nonlinearity. For these models, it is assumed that the bridge deck and girders behave elastically with no damage. This assumption is consistent with past studies and post-earthquake inspections [5,15]. Nielson and DesRoches [5] modeled the bridge girders and deck as a single beam with stiffness properties determined from the composite multi-girder and deck section. This study employed a more detailed grid of beam elements to better model the vertically and horizontally distributed stiffness and mass of the girder and deck system, similar to the grid model described in Filipov et al. [15].

For columns, flexural and/or combined axial-flexural damage are commonly observed in earthquakes; however, columns in low-seismic regions may also be more susceptible to shear failure modes because of poor confinement and shear reinforcement found in non-seismically detailed columns. This study uses a rotational spring at the tops and bottoms of the bent columns to simulate flexure and shear failure in columns with and without seismic detailing. The nonlinear behavior of the rotational springs was calibrated based on a large database of 319 and 171 rectangular and circular columns, respectively. These columns were tested under cyclic loading with various levels of seismic detailing and shear reinforcement [16,17]. The nonlinear springs also reflect ACI [18] backbone strength parameters for flexure, lap splice, and shear failure modes; see elsewhere for details [12].

Moreover, steel girder bridges available in Texas typically have steel fixed and expansion bearing types, where a fixed bearing allows only rotational movement (no translation) and an expansion bearing allows both rotation and horizontal longitudinal translation. For such bearings, nonlinear models under lateral loads are developed and calibrated with extensive experimental data available in the research literature [19,20].

For other components of the bridge models, including expansion joints, deck pounding, abutments, and foundations, nonlinear models developed in previous studies [5,15,21], with appropriate modifications to represent details typical of Texas, are also assigned. It should be noted that from review of as-built drawings from Texas bridges, it was observed that most steel girder bridges in Texas have pile-bent abutments, which have two types of resistance in the longitudinal direction: passive resistance and active resistance. Passive resistance is developed as a result of pressing the abutment into the soil. In this case, both soil and piles beneath the abutment provide a contribution. Active resistance is developed as a result of pulling away the abutment from the backfill. In this case, the resistance is only provided by piles beneath the abutment. For the transverse direction, only the piles contribute to resistance.

The nonlinear 3-D model of each sampled bridge is subjected to 10 randomly selected ground motions that are scaled to different values of PGA, which leads to 640 nonlinear time history analyses. The analyses are conducted using the DesignSafe Cyberinfrastructure [22]. DesignSafe is a new cyberinfrastructure for natural hazards engineering that is able to perform numerical simulations using high-performance computing. In this study, it is assumed that damage can occur in columns, bearings, and abutments; therefore, the responses of these components are recorded during each analysis. In particular, as for column response, the rotational ductility, defined as the maximum column rotation divided by the yield rotation, is captured. For bearings, the longitudinal and transverse deformations of both fixed and expansion bearings are recorded during the analyses. Finally, for abutments, deformations in passive, active, and transverse directions are documented. These outputs are set as inputs for the probabilistic seismic demand model, which is presented in the next section.

## Probabilistic Seismic Demand Model

Given the output of a nonlinear time history analysis, Cornel, et al. [6] showed that the median of seismic demands follows a power function of intensity measure as follows:

$$S_D = a\text{PGA}^b \tag{3}$$

This equation can be rearranged to logarithm space where $\ln(S_D)$ follows a linear function of PGA with coefficients $\ln(a)$ and $b$. Therefore, coefficients $a$ and $b$ can be computed by fitting a linear regression to the lognormal of the outputs from nonlinear time history analyses. Moreover, Cornell et al. [6] proposed that the conditional seismic demands typically follow a lognormal distribution, resulting in normal distribution with median of $\ln(S_D)$ and dispersion of $\beta_{D|PGA}$, in the transformed space. The variation or dispersion of the seismic demands about the mean, given the intensity measure, is the conditional lognormal standard deviation of the seismic demand ($\beta_{D|PGA}$). According to Padgett et al. [23], $\beta_{D|PGA}$ is approximately estimated by computing the dispersion of the data around the fitted linear regression using the following equation:

$$\beta_{D|IM} = \sqrt{\frac{\sum_{i=1}^{N}[\ln(d_i) - \ln(S_D)]^2}{N-2}} \tag{4}$$

Recall that in this study, the seismic demands of columns, bearings, and abutments are recorded. In particular, the seismic demands comprise the rotational ductility of columns, the longitudinal and transverse deformations of both fixed and expansion bearings, and deformations in passive, active, and transverse directions of abutments. The parameters of PSDM for the mentioned seismic demands are shown in Table 2.

Table 2. Probabilistic seismic demand parameter estimations

| Component | Abbreviation | $a$ | $b$ | $\beta_{D|PGA}$ |
|---|---|---|---|---|
| Column | $\mu_\theta$ | 0.83 | 0.97 | 0.86 |
| Fixed bearing-Long. | fx_L | 0.68 | 1.18 | 1.04 |
| Fixed bearing-Trans. | fx_T | 0.54 | 2.14 | 1.47 |
| Expan. bearing-Long. | ex_L | 1.16 | 1.07 | 0.79 |
| Expan. bearing-Trans. | ex_T | 1.11 | 1.37 | 0.93 |
| Abutment-Active | abut_A | 0.05 | 0.61 | 1.13 |
| Abutment-Passive | abut_P | 0.10 | 0.87 | 1.80 |
| Abutment-Trans. | abut_T | 0.02 | 0.14 | 0.91 |

## Probabilistic Seismic Capacity Model

For each component, four levels of damage, i.e., limit states, are defined as Slight, Moderate, Excessive, and Complete. According to past studies [6,13], it is assumed that the capacity for each limit state follows a lognormal distribution with a median of $S_C$ and dispersion of $\beta_C$. The values of $S_C$ are assumed based on engineering judgement and test results. Table 3 shows the component capacities used in this study for different limit states. For more information about the details of the considered component capacities, readers are referred to Khosravikia et al. [12].

As an example, the limit states suggested for reinforced concrete column are briefly discussed here. In this study, the limit states suggested by Ramanathan [24] for strength-degrading columns, typical of California bridges from the 1970s and 1980s, are used. This assumption is owing to the fact that the reinforced concrete columns found in Texas have similar details as the strength-degrading columns found in the Ramanathan [24] study; see elsewhere for details [12]. Ramanathan [24] proposed median curvature ductilities of 1.0, 2.0, 3.5, and 5.0 for Slight, Moderate, Extensive, and Complete limit states, respectively, which are respectively associated with the points that the column starts yielding, cracking, spalling, and buckling. It is worth noting that the median values suggested by Ramanathan [24] are curvature ductilities, and they should be translated into equivalent rotational ductilities. To do so, the conversion method suggested by Khosravikia et al. [12] in used, which results in median rotational ductilities of 1.0, 2.5, 3.25, and 4.0 for Slight, Moderate, Extensive, and Complete limit states, respectively.

Moreover, to account for the uncertainty in capacity of each component, the coefficient of variation (COV) of 25% for slight and moderate, and 50% for extensive and complete limit states are taken into account, which results in $\beta_C$ of 0.25 for slight and moderate, and $\beta_C$ of 0.47 for extensive and complete limit states using the following equation:

$$\beta_C = \sqrt{1+\text{COV}^2} \tag{5}$$

Table 3. Component capacities

| Component | | Slight | | Moderate | | Extensive | | Complete | |
|---|---|---|---|---|---|---|---|---|---|
| | | $S_C$ | $\beta_C$ | $S_C$ | $\beta_C$ | $S_C$ | $\beta_C$ | $S_C$ | $\beta_C$ |
| Column | $\mu_\theta$ | 1 | 0.25 | 2.5 | 0.25 | 3.25 | 0.47 | 4 | 0.47 |
| Steel Fixed bearing-Long. (in) | fx_L | 0.25 | 0.25 | 0.75 | 0.25 | 1.5 | 0.47 | 7.25 | 0.47 |
| Steel Fixed bearing-Trans. (in) | fx_T | 0.25 | 0.25 | 0.75 | 0.25 | 1.5 | 0.47 | 7.25 | 0.47 |
| Steel Expan. bearing-Long. (in) | ex_L | 1.5 | 0.25 | 3.5 | 0.25 | 5 | 0.47 | 7.25 | 0.47 |
| Steel Expan. bearing-Trans. (in) | ex_T | 0.25 | 0.25 | 0.75 | 0.25 | 1.5 | 0.47 | 7.25 | 0.47 |
| Abutment-Passive (in) | abut_P | 1.25 | 0.25 | 6.0 | 0.25 | 8.0 | 0.47 | 10.0 | 0.47 |
| Abutment-Active (in) | abut_A | 0.375 | 0.25 | 1.5 | 0.25 | 3 | 0.47 | 8 | 0.47 |
| Abutment-Trans (in) | abut_T | 0.375 | 0.25 | 1.5 | 0.25 | 3 | 0.47 | 8 | 0.47 |

**Component and Bridge Fragility Curves**

Given demand and capacity models for each component, the probability of damage can be computed using Eq. 2. Figure 5 shows the fragility curves for different components. Each plot represents the fragility functions of the components for one specific limit state (e.g., slight, moderate, extensive, and complete). As seen, regardless of the limit state, bearings are the most vulnerable components and are most likely to experience damage in an earthquake. This is mainly due to the fact that steel girder bridges, based on the age of the bridge samples, have steel bearings which are known to be vulnerable to seismic hazards. Although this type of bearings has recently been replaced by elastomeric pads in modern steel girder bridges, the majority of the steel bridges constructed in Texas have steel bearings [12]. Moreover, for expansion bearings, there is a significant difference in the fragility curves for longitudinal and transverse directions. In particular, this type of bearing is more vulnerable in the transverse direction. This observation is mainly because of the fact that this type of bearings, unlike fixed bearings, allows horizontal longitudinal translation. It is also found that abutments are the least vulnerable component of the bridges. Due to their relative flexibility and lower strength, the bearings are expected to experience large deformations and subsequent damage, limiting the loads that can be transferred to the abutments. These fragility curves can be used as a guidance for post-event bridge inspection, to identify the critical components most likely to exhibit damage.

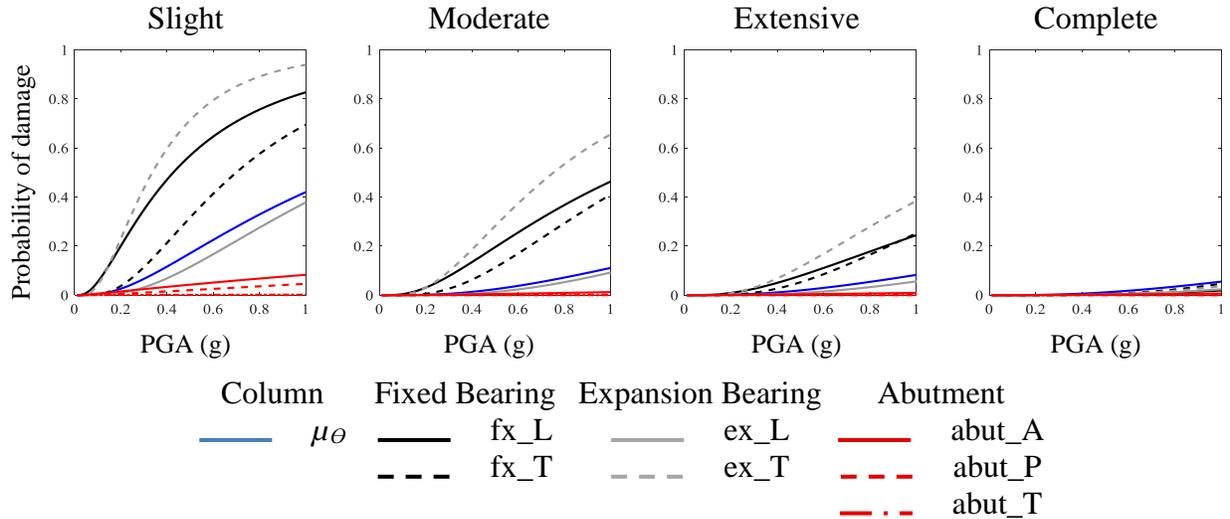

Figure 5. Component fragility curves

To compute the fragility curves for the bridges considered in this study, it is assumed that if any component exceeds any level of damage, then the whole bridge exceeds that level of damage. It is a conservative assumption, though it is consistent with other past research [25]. In fact, for each level of damage, it is assumed that the probability of damage for a bridge is the union of the probabilities that bridge components exceed that limit state, given mathematically as:

$$P(\text{Fail}_{system} | \text{LS}) = \bigcup_{i=1}^{n} P(\text{Fail}_{componenet-i} | \text{LS}) \tag{6}$$

This study utilizes Monte Carlo sampling analysis to compute the above-mentioned probability. In particular, for each PGA, $10^6$ random samples are generated for both demand and capacity sides of the components based on their demand and capacity models. It should be noted that the correlation of the seismic demands in various components are taken into account when random realizations are generated; see elsewhere for details [12]. For each pre-specified level of damage, the randomly generated demand and capacity realizations for each component are compared to see if the demand of the component exceeds its specified level of damage for that component. In each sample of Monte Carlo analysis, if any component exceeds the specific level of damage, then the whole bridge exceeds that level of damage. After $10^6$ samples, the probability of damage for an entire bridge is computed for a given PGA. This sampling is then carried out over a wide range of PGA values to compute the probability of damage for different values of PGA, which is the underlying data for the generation of bridge fragility curves. Fragility curves are assumed to follow a lognormal distribution and are, therefore, produced via a simple linear regression of the underlying failure probabilities at various PGA levels to estimate the median PGA (i.e., PGA corresponding to 50% likelihood of exceeding the specified limit state) and dispersion of the fragility functions. Table 4 presents the median PGA and dispersion values for the system-level fragility curves of steel girder bridges. For instance, the median of exceeding the slight limit state is 0.27 g, which means that for ground motions with PGA of 0.27 g, there is 50% probability of exceeding slight damage for steel girder bridges.

Table 4. Median PGA and dispersion of fragility curves

| Limit State | Median (g) | Stdv |
|---|---|---|
| Slight | 0.27 | 0.72 |
| Moderate | 0.64 | 0.72 |
| Extensive | 0.99 | 0.75 |
| Complete | 2.69 | 0.83 |

Figure 6 presents the fragility curves for steel girder bridges based on the median and dispersion values shown in Table 4. It should be noted that the largest PGA values recorded in the ground motions used in this study came from the November 7, 2016 Cushing, Oklahoma M5.0 event. The largest PGA recorded during this event was approximately 0.59g, which was at a hypocentral distance of 5.2km [12]. Other stations ranging from 6.4km to 9.6km from the hypocenter recorded peak PGA values ranging from 0.20g to 0.32g. These data suggest that larger magnitude induced earthquake (M5+) have a significant likelihood of producing slight to moderate damage in steel girder bridges nearby the hypocenter (e.g., hypocentral distances less than approximately 10km). The developed fragility curves can be used to inform post-earthquake inspection decisions for this bridge class. For example, key stakeholder can use the fragility curves to identify threshold PGA values that produce a sufficient likelihood of damage warranting closure and/or inspection. This threshold PGA, coupled with up-to-the-minute ground shaking estimations from the USGS ShakeMaps can be used to identify the geographic area over which action must be taken for this bridge class.

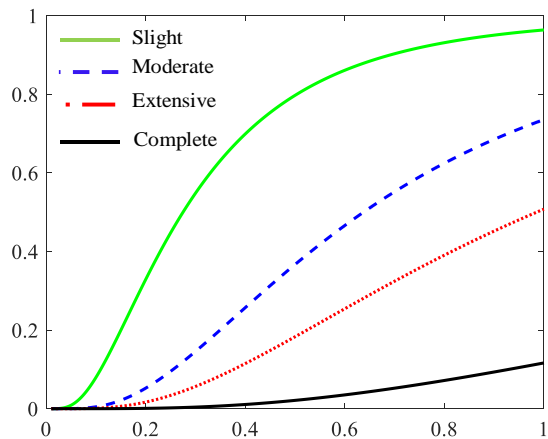

Figure 6. Fragility curves of steel girder bridges

## Conclusion

This study evaluates the vulnerability of highway bridges in areas with potentially-induced hazards largely associated with petroleum and gas activities. Such earthquakes can occur in areas that historically have had negligible seismicity. The recent increase in the rate of such earthquakes raises the question over the seismic performance of the infrastructure that was designed without considering seismic demands. The case study in this paper focuses on multi span, simply supported steel girder bridges located in Texas, which is mainly believed to be subjected to induced seismic hazards. The assessment in this study was conducted by developing fragility curves using 3-D bridge models, a suite of ground motions created for Texas, and nonlinear time-history analyses. The results uncover that that potentially-induced earthquakes are capable of causing damage in steel girder bridges, particularly for some of the largest magnitude induced events that have been recently recorded. Ongoing research is currently being conducted by the authors to evaluate the effect of induced hazards on different types of infrastructure, including other bridge and residential infrastructure. This new fragility information will be used to inform how induced seismicity should be considered in seismic risk assessments in the Central and Eastern United States.

## Acknowledgments

The financial support from the Texas Department of Transportation (TxDOT) through Grant Number 0-6916 is gratefully acknowledged. The opinions and findings expressed herein are those of the authors and not the sponsors